\documentclass[italian,english]{article}
\usepackage[latin9]{inputenc}
\usepackage{color}
\usepackage{amsmath}
\usepackage{amssymb}

\newcommand{\lyxaddress}[1]{
\par {\raggedright #1
\vspace{1.4em}
\noindent\par}
}

\usepackage{babel}

\begin{document}

\title{\textbf{Gravitational waves astronomy: the ultimate test for Einstein's
General Relativity }}

\author{\textbf{Christian Corda}}

\maketitle

\lyxaddress{\begin{center}
Associazione Scientifica Galileo Galilei, Via Pier Cironi 16 - 59100
PRATO, Italy 
\par\end{center}}

\begin{center}
\textit{E-mail address:} \textcolor{blue}{cordac.galilei@gmail.com} 
\par\end{center}
\begin{abstract}
It is well known that Einstein's General Relativity (GR) achieved
a great success and overcame lots of experimental tests. On the other
hand, GR also showed some shortcomings and flaws which today advise
theorists to ask if it is the definitive theory of gravity. In this
review we show that, if advanced projects on the detection of Gravitational
Waves (GWs) will improve their sensitivity, allowing to perform a
GWs astronomy, understanding if Einstein's GR is the correct and definitive
theory of gravity will be possible. For this goal, accurate angular
and frequency dependent response functions of interferometers for
GWs arising from various Theories of Gravity, i.e. GR and Extended
Theories of Gravity will have to be used. 

This review is founded on the Essay which won an Honorable Mention
at the the 2009 Gravity Research Foundation Awards.
\end{abstract}
The scientific community aims in a first direct detection of GWs in
next years (for the current status of GWs interferometers see \cite{key-1})
confirming the indirect, Nobel Prize Winner, proof of Hulse and Taylor
\cite{key-2}. 

Detectors for GWs will be important for a better knowledge of the
Universe and either to confirm or rule out the physical consistency
of GR or of any other theory of gravitation \cite{key-3,key-4,key-5,key-6,key-7,key-8,key-9}.
In fact, in the context of Extended Theories of Gravity, some differences
between GR and the others theories can be pointed out starting by
the linearized theory of gravity \cite{key-3,key-4,key-5,key-6,key-7,key-8,key-9,key-10,key-11,key-12,key-13,key-14}.
In this picture, detectors for GWs are in principle sensitive also
to a hypothetical \textit{scalar} component of gravitational radiation,
that appears in extended theories of gravity like scalar-tensor gravity
\cite{key-5,key-11,key-13,key-14}, bi-metric theory \cite{key-6},
high order theories \cite{key-3,key-4,key-7,key-8,key-9,key-10,key-12},
Brans-Dicke theory \cite{key-15} and string theory \cite{key-16}.

Motivations on an potential extension of GR arise from the fact that,
even if Einstein's Theory \cite{key-17} achieved a great success
(see for example the opinions and of Wheeler who says that {}``\textit{Among
all bodies of physical laws none has ever been found that is simpler
or more beautiful than Einstein's geometric theory of gravity}''
\cite{key-17} and of Landau who says that {}``\textit{General Relativity
is, together with Quantum Field Theory, the best scientific theory
of all}'' \cite{key-18}) and overcame lots of experimental tests
\cite{key-17}, it also showed some shortcomings and flaws \cite{key-19,key-20,key-21}.
Thus, today theorists ask if it is the correct and ultimate theory
of gravity. On the other hand, GR is very difficult to be quantized.
This point makes GR different from other field theories like the electromagnetic
theory, and also rules out the possibility of treating gravitation
like other quantum theories, precluding the unification of gravity
with other interactions. At the present time, a consistent Quantum
Gravity Theory which leads to the unification of gravitation with
the other forces has not been realized \cite{key-20,key-21}. 

Another point of view defines \textit{Extended Theories of Gravity}
those semiclassical theories where the Lagrangian is modified, in
respect to the standard Einstein-Hilbert gravitational Lagrangian,
adding high-order terms in the curvature invariants (terms like $R^{2}$,
$R^{\alpha\beta}R_{\alpha\beta}$, $R^{\alpha\beta\gamma\delta}R_{\alpha\beta\gamma\delta}$,
$R\Box R$, $R\Box^{k}R$) or terms with scalar fields non minimally
coupled to geometry (terms like $\phi^{2}R$) \cite{key-19,key-20,key-21}.
Terms like those are present, in general, in all the approaches to
perform the unification between gravity and other interactions. It
is also important to stress that, from a cosmological point of view,
these modifies of GR generate inflationary frameworks which solve
lots of problems of the Standard Universe Model \cite{key-22,key-23,key-24}.
Notice that we are not telling that GR is wrong. We are sure that,
even in the context of Extended Theories, GR remains the most important
part of the structure \cite{key-3,key-4,key-5,key-6,key-7,key-8,key-9,key-10,key-11,key-18,key-19,key-20,key-21}.
We only would like to understand if weak modifies on GR's structure
could be needed to solve some theoretical and observing problems \cite{key-19,key-20,key-21}.
In this tapestry, even Einstein told that General Relativity could
not be definitive \cite{key-25} as during his famous research on
the Unified Field Theory, he tried to realize a theory that he called
{}``Generalized Theory of Gravitation'', and he said that  mathematical
difficulties precluded him to derive the final equations \cite{key-25}.
On the other hand, it is well known that various Extended Theories
of gravity are banned by requirements of cosmology and Solar System
tests and, in general, the modification in respect to standard GR
has to be very weak in order to satisfy such constrains \cite{key-26,key-27}.

In the general framework of cosmological evidences, other considerations
suggest an extension of GR. In fact, the accelerated expansion of
the Universe, which is today observed, shows that cosmic dynamic is
dominated by the so called Dark Energy, which gives a large negative
pressure. In this standard picture, such new ingredient is considered
as a source of the \textit{right side} of the field equations. It
could be some form of un-clustered non-zero vacuum energy which, together
with the clustered Dark Matter, drives the global dynamics. This is
the famous {}``concordance model'' ($\Lambda$CDM) which gives a
good tapestry of the today observed Universe. However, even if in
agreement with the CMBR, LSS and SNeIa data, such a model presents
several shortcomings as the well known {}``coincidence'' and {}``cosmological
constant'' problems \cite{key-28}. The alternative approach proposed
by theorists changes the \textit{left side} of the field equations,
seeing if observed cosmic dynamics can be achieved extending GR \cite{key-19,key-20,key-21,key-29}.
In this case, it is not required to find out candidates for Dark Energy
and Dark Matter, that, till now, have not been found, but only the
{}``observed'' ingredients, which are curvature and baryon matter,
have to be taken into account. From this point of view, gravity could
be different at various scales \cite{key-29} and alternative theories
can be considered. By thinking in this way, the most popular Dark
Energy and Dark Matter models can be achieved considering $f(R)$
theories of gravity, where $R$ is the Ricci curvature scalar, and/or
Scalar-Tensor Gravity \cite{key-19,key-20,key-21}.

The aim of this review is showing that, if advanced projects on the
detection of GWs will improve their sensitivity, allowing to perform
a GWs astronomy \cite{key-1}, it will be ultimately possible to understand
if Einstein's GR is the correct and definitive theory of gravity \cite{key-3}.
For this goal, accurate angular and frequency dependent response functions
of interferometers for GWs arising from various Theories of Gravity,
i.e. GR and Extended Theories of Gravity will have to be used \cite{key-3,key-4,key-5,key-6,key-8,key-9,key-10,key-30,key-31}.
The papers which found this review paper have been the world's most
cited within the official Astroparticle Publication Review of ASPERA
during the 2007 with 13 citations \cite{key-32}. ASPERA is the network
of national government agencies responsible for coordinating and funding
national research efforts in Astroparticle Physics, see \cite{key-32}.
This review is founded on the essay which won an Honorable Mention
at the the 2009 Gravity Research Foundation Awards \cite{key-3}.

Working with $G=1$, $c=1$ and $\hbar=1$ (natural units), the line
element for a GW arising from standard General Relativity and propagating
in the $z$ direction is \cite{key-3,key-17,key-30,key-31,key-33,key-34,key-35} 

\begin{equation}
ds^{2}=dt^{2}-dz^{2}-(1+h_{+})dx^{2}-(1-h_{+})dy^{2}-2h_{\times}dxdy,\label{eq: metrica TT totale}\end{equation}

where $h_{+}(t+z)$ and $h_{\times}(t+z)$ are the weak perturbations
due to the $+$ and the $\times$ polarizations which are expressed
in terms of synchronous coordinates in the Transverse Traceless (TT)
gauge \cite{key-17}. The total frequency and angular dependent response
function (i.e. the detector pattern) to the $+$ polarization of an
interferometer with arms in the $u$ and $v$ directions in respect
to the propagating GW has been computed in \cite{key-3,key-30,key-31},
it is:\begin{align}
\tilde{H}^{+}(\omega) & \equiv\Upsilon_{u}^{+}(\omega)-\Upsilon_{v}^{+}(\omega)\nonumber \\
 & =\frac{(\cos^{2}\theta\cos^{2}\phi-\sin^{2}\phi)}{2L}\tilde{H}_{u}(\omega,\theta,\phi)-\frac{(\cos^{2}\theta\sin^{2}\phi-\cos^{2}\phi)}{2L}\tilde{H}_{v}(\omega,\theta,\phi),\label{eq: risposta totale Virgo +}\end{align}

that, in the low frequencies limit ($\omega\rightarrow0$) gives the
well known low frequency response function of \cite{key-33,key-34}
for the $+$ polarization: 

\begin{equation}
\tilde{H}^{+}(\omega)=\frac{1}{2}(1+\cos^{2}\theta)\cos2\phi+O\left(\omega\right)\,.\label{eq: risposta totale approssimata}\end{equation}

Sketching the derivation of eq. (\ref{eq: risposta totale Virgo +})
is important for a sake of clearness \cite{key-3}.

Following \cite{key-3,key-30,key-31}, the rotation in respect to
the $u$ and $v$ directions is

\begin{equation}
\begin{array}{ccc}
x & = & -u\cos\theta\cos\phi-v\cos\theta\sin\phi+w\sin\theta\\
\\y & = & u\sin\phi-v\cos\phi\\
\\z & = & u\sin\theta\cos\phi+v\sin\theta\sin\phi+w\cos\theta.\end{array}\label{eq: rotazione 2}\end{equation}

Then, the line element transforms as \cite{key-3,key-30,key-31}

\begin{equation}
g^{ik}=\frac{\partial x^{i}}{\partial x'^{l}}\frac{\partial x^{k}}{\partial x'^{m}}g'^{lm}.\label{eq: trasformazione metrica}\end{equation}

Taking into account only the $+$ polarization and using eqs. (\ref{eq: rotazione 2})
and (\ref{eq: trasformazione metrica}), the line element in the $\overrightarrow{u}$
direction becomes:

\begin{equation}
ds^{2}=-dt^{2}+[1+(\cos^{2}\theta\cos^{2}\phi-\sin^{2}\phi)h_{+}(t+u\sin\theta\cos\phi)]du^{2}.\label{eq: metrica + lungo u}\end{equation}

A good way to analyse variations in the proper distance (time) is
by means of {}``bouncing photons'' \cite{key-3,key-30,key-31}.
A photon can be launched from the interferometer's beam-splitter to
be bounced back by the mirror. This kind of analysis was created by
Rakhmanov in \cite{key-35}. Actually, it has been strongly generalized
to angular dependences, scalar waves and massive GWs in \cite{key-3,key-4,key-5,key-9,key-13,key-30,key-31}.

The condition for null geodesics ($ds^{2}=0$) in eq. (\ref{eq: metrica + lungo u})
gives the coordinate velocity of the photon:

\begin{equation}
v_{p}^{2}\equiv(\frac{du}{dt})^{2}=\frac{1}{[1+(\cos^{2}\theta\cos^{2}\phi-\sin^{2}\phi)h_{+}(t+u\sin\theta\cos\phi)]},\label{eq: velocita' fotone u}\end{equation}

which will be used for calculations of the photon propagation time
between the beam-splitter and the mirror \cite{key-3,key-30,key-31,key-35}.
If one assumes that the beam splitter is located in the origin of
the new coordinate system (i.e. $u_{b}=0$, $v_{b}=0$, $w_{b}=0$)
the analysis is simplified. As we are in the TT gauge, the coordinates
of the beam-splitter $u_{b}=0$ and of the mirror $u_{m}=L$ do not
change under the influence of the GW \cite{key-3,key-17,key-30,key-31,key-35},
thus the duration of the forward trip is given by

\begin{equation}
T_{1}(t)=\int_{0}^{L}\frac{du}{v_{p}(t'+u\sin\theta\cos\phi)},\label{eq: durata volo}\end{equation}

with 

\begin{center}
$t'=t-(L-u)$.
\par\end{center}

In the equation (\ref{eq: durata volo}) $t'$ is the delay time (i.e.
$t$ is the time at which the photon arrives in the position $L$,
so $L-u=t-t'$).

At first order in $h_{+}$ the integral (\ref{eq: durata volo}) can
be approximated with

\begin{equation}
T_{1}(t)=T+\frac{\cos^{2}\theta\cos^{2}\phi-\sin^{2}\phi}{2}\int_{0}^{L}h_{+}(t'+u\sin\theta\cos\phi)du,\label{eq: durata volo andata approssimata u}\end{equation}

where $T=L$ (recall that natural units are used) is the transit time
of the photon in absence of the GW. Similarly, the duration of the
return trip is\begin{equation}
T_{2}(t)=T+\frac{\cos^{2}\theta\cos^{2}\phi-\sin^{2}\phi}{2}\int_{L}^{0}h_{+}(t'+u\sin\theta\cos\phi)(-du),\label{eq: durata volo ritorno approssimata u}\end{equation}

and now the delay time is 

\begin{center}
$t'=t-(u-l)$.
\par\end{center}

The round-trip time is the sum of $T_{2}(t)$ and $T_{1}[t-T_{2}(t)]$.
As the difference between the exact and the approximate values is
second order in $h_{+}$, $T_{1}[t-T_{2}(t)]$ can be approximated
by $T_{1}(t-T)$. Then, to first order in $h_{+}$, the duration of
the round-trip is

\begin{equation}
T_{r.t.}(t)=T_{1}(t-T)+T_{2}(t).\label{eq: durata round trip}\end{equation}

By using eqs. (\ref{eq: durata volo andata approssimata u}) and (\ref{eq: durata volo ritorno approssimata u})
one gets that deviations of this round-trip time (i.e. proper distance)
from its unperturbed value are given by

\begin{equation}
\begin{array}{c}
\delta T(t)=\frac{\cos^{2}\theta\cos^{2}\phi-\sin^{2}\phi}{2}\int_{0}^{L}[h_{+}(t-2T-u(1-\sin\theta\cos\phi))+\\
\\+h_{+}(t+u(1+\sin\theta\cos\phi))]du.\end{array}\label{eq: variazione temporale in u}\end{equation}

Introducing the Fourier transform of the $+$ polarization of the
field, defined by

\begin{equation}
\tilde{h}_{+}(\omega)\equiv\int_{-\infty}^{\infty}dth_{+}(t)\exp(i\omega t),\label{eq: TF}\end{equation}

and using the Fourier translation theorem, in the frequency domain
it is:

\begin{equation}
\delta\tilde{T}(\omega)=\frac{1}{2}(\cos^{2}\theta\cos^{2}\phi-\sin^{2}\phi)\tilde{H}_{u}(\omega,\theta,\phi)\tilde{h}_{+}(\omega),\label{eq: segnale in frequenza lungo u}\end{equation}

where

\begin{equation}
\begin{array}{c}
\tilde{H}_{u}(\omega,\theta,\phi)=\frac{-1+\exp(2i\omega L)}{2i\omega(1+\sin^{2}\theta\cos^{2}\phi)}+\\
\\+\frac{-\sin\theta\cos\phi((1+\exp(2i\omega L)-2\exp i\omega L(1-\sin\theta\cos\phi)))}{2i\omega(1+\sin\theta\cos^{2}\phi)}\end{array}\label{eq: fefinizione Hu}\end{equation}

and $\tilde{H}_{u}(\omega,\theta,\phi)\rightarrow L$ when $\omega\rightarrow0$.

Thus, if one defines a {}``signal'' in the $u$ arm like $S(\omega)\equiv\frac{\delta\tilde{T}(\omega)}{2T},$
the total response function of this arm of the interferometer to the
$+$ component is:

\begin{equation}
\Upsilon_{u}^{+}(\omega)\equiv\frac{S(\omega)}{\tilde{h}_{+}(\omega)}=\frac{(\cos^{2}\theta\cos^{2}\phi-\sin^{2}\phi)}{2L}\tilde{H}_{u}(\omega,\theta,\phi).\label{eq: risposta + lungo u}\end{equation}

In the same way, one gets the response function of the $v$ arm of
the interferometer to the $+$ polarization: 

\begin{equation}
\Upsilon_{v}^{+}(\omega)=\frac{(\cos^{2}\theta\sin^{2}\phi-\cos^{2}\phi)}{2L}\tilde{H}_{v}(\omega,\theta,\phi)\label{eq: risposta + lungo v}\end{equation}

where, now 

\begin{equation}
\begin{array}{c}
\tilde{H}_{v}(\omega,\theta,\phi)=\frac{-1+\exp(2i\omega L)}{2i\omega(1+\sin^{2}\theta\sin^{2}\phi)}+\\
\\+\frac{-\sin\theta\sin\phi((1+\exp(2i\omega L)-2\exp i\omega L(1-\sin\theta\sin\phi)))}{2i\omega(1+\sin^{2}\theta\sin^{2}\phi)},\end{array}\label{eq: fefinizione Hv}\end{equation}

with $\tilde{H}_{v}(\omega,\theta,\phi)\rightarrow L$ when $\omega\rightarrow0$. 

The total response function is the difference between (\ref{eq: risposta + lungo u})
and (\ref{eq: risposta + lungo v}), thus one obtains immediately
eq. (\ref{eq: risposta totale Virgo +}).

The same analysis works for the $\times$ polarization (see \cite{key-30,key-31}
for details). At the end, the total frequency and angular dependent
response function of an interferometer to the $\times$ polarization
is:

\begin{equation}
\tilde{H}^{\times}(\omega)=\frac{-\cos\theta\cos\phi\sin\phi}{L}[\tilde{H}_{u}(\omega,\theta,\phi)+\tilde{H}_{v}(\omega,\theta,\phi)],\label{eq: risposta totale Virgo per}\end{equation}
that, in the low frequencies limit ($\omega\rightarrow0$), gives
the low frequency response function of \cite{key-33,key-34} for the
$\times$ polarization: \begin{equation}
\tilde{H}^{\times}(\omega)=-\cos\theta\sin2\phi+O\left(\omega\right)\,.\label{eq: risposta totale approssimata 2}\end{equation}

The case of massless Scalar-Tensor Gravity has been discussed in \cite{key-5,key-13}
with a {}``bouncing photons analysis'' similar to the previous one.
In this case, the line-element in the TT gauge can be extended with
one more polarization, labelled with $\Phi(t+z)$, i.e.

\begin{equation}
ds^{2}=dt^{2}-dz^{2}-(1+h_{+}+\Phi)dx^{2}-(1-h_{+}+\Phi)dy^{2}-2h_{\times}dxdy.\label{eq: metrica TT super totale}\end{equation}

The total frequency and angular dependent response function of an
interferometer to this {}``scalar'' polarization is \cite{key-5,key-13}\begin{align}
\tilde{H}^{\Phi}(\omega) & =\frac{\sin\theta}{2i\omega L}\{\cos\phi[1+\exp(2i\omega L)-2\exp i\omega L(1+\sin\theta\cos\phi)]+\nonumber \\
 & -\sin\phi[1+\exp(2i\omega L)-2\exp i\omega L(1+\sin\theta\sin\phi)]\}\,,\label{eq: risposta totale Virgo scalar}\end{align}

that, in the low frequencies limit ($\omega\rightarrow0$), gives
the low frequency response function of \cite{key-16} for the $\Phi$
polarization: \textbf{\begin{equation}
\tilde{H}^{\Phi}(\omega)=-\sin^{2}\theta\cos2\phi+O(\omega).\label{eq: risposta totale approssimata scalar}\end{equation}
}

In \cite{key-13} the response function (\ref{eq: risposta totale Virgo scalar})
has been used to study the cross-correlation between the Virgo interferometer
and the monopole mode of the MiniGRAIL resonant sphere for the detection
of massless stochastic scalar GWs. Even if such a cross correlation
is very small, a maximum is present at about $2710Hz$, i.e. within
the sensitivity's range of both of MiniGRAIL and Virgo \cite{key-13}.
Then, if the eventual detection of a monopole mode of the MiniGRAIL
bar at about $2710Hz$ will coincide with a signal detected by the
Virgo interferometer at the same frequency, such a detection will
be a strong endorsement for massless Scalar Tensor Gravity. Indeed,
the monopole mode of a sphere cannot be excited by ordinary tensor
waves arising from standard GR, see \cite{key-13} for details.

The cases of massive Scalar-Tensor Gravity and $f(R)$ theories are
totally equivalent \cite{key-3,key-4,key-5,key-8}. This is not a
surprise as it is well known that there is a more general conformal
equivalence between Scalar-Tensor Gravity and $f(R)$ theories, even
if there is a large debate on the possibility that such a conformal
equivalence should be a \emph{physical} equivalence too \cite{key-19,key-20,key-21}.
In such cases, the presence of a small mass generates a longitudinal
component in the third polarization. Thus, the extension of the TT
gauge to the third massive mode is impossible \cite{key-3,key-4,key-5,key-8}.
But gauge transformations permit to write the line-element due to
such a third scalar mode in a conformally flat form \cite{key-3,key-4,key-5,key-8}: 

\begin{equation}
ds^{2}=[1+\Phi(t-v_{G}z)](-dt^{2}+dz^{2}+dx^{2}+dy^{2}).\label{eq: metrica puramente scalare}\end{equation}

Assuming that the interferometer arm is parallel to the propagating
GW, a longitudinal response function can be associated to such a massive
mode \cite{key-3,key-4,key-5,key-8}: \begin{equation}
\begin{array}{c}
\Upsilon_{l}(\omega)=\frac{1}{m^{4}\omega^{2}L}(\frac{1}{2}(1+\exp[2i\omega L])m^{2}\omega^{2}L(m^{2}-2\omega^{2})+\\
\\-i\exp[2i\omega L]\omega^{2}\sqrt{-m^{2}+\omega^{2}}(4\omega^{2}+m^{2}(-1-iL\omega))+\\
\\+\omega^{2}\sqrt{-m^{2}+\omega^{2}}(-4i\omega^{2}+m^{2}(i+\omega L))+\\
\\+\exp[iL(\omega+\sqrt{-m^{2}+\omega^{2}})](m^{6}L+m^{4}\omega^{2}L+8i\omega^{4}\sqrt{-m^{2}+\omega^{2}}+\\
\\+m^{2}(-2L\omega^{4}-2i\omega^{2}\sqrt{-m^{2}+\omega^{2}}))+2\exp[i\omega L]\omega^{3}(-3m^{2}+4\omega^{2})\sin[\omega L]).\end{array}\label{eq: risposta totale lungo z massa}\end{equation}

Eq. (\ref{eq: risposta totale lungo z massa}) has been obtained in
\cite{key-4} with the {}``bouncing photons analysis'' and in \cite{key-8}
with a different treatment that used geodesic deviation. $m$ in eq.
(\ref{eq: risposta totale lungo z massa}) is the small mass of the
particle associated to the GW and $v_{G}$ in eq. (\ref{eq: metrica puramente scalare})
is the particle's velocity. In fact, the group velocity can be expressed
in terms of a wave-packet \cite{key-4,key-8}. In this case, the relation
mass-velocity is $m=\sqrt{(1-v_{G}^{2})}\omega,$ see \cite{key-4,key-8}
for details.

As signals from GWs are quite weak \cite{key-1}, in order to discriminate
between various signals, advanced projects on the detection of GWs
will have to improve their sensitivity allowing to perform a GWs astronomy
\cite{key-3}. Then, one will only have to look the interferometer
response functions to understand if GR is the ultimate theory of gravity.
If only the two response functions (\ref{eq: risposta totale Virgo +})
and (\ref{eq: risposta totale Virgo per}) will be present, the conclusion
will be that GR is definitive. If the response function (\ref{eq: risposta totale Virgo scalar})
will be present too, then massless Scalar - Tensor Gravity will be
the correct theory of gravitation. Finally, if a longitudinal response
function will be present, i.e. Eq. (\ref{eq: risposta totale lungo z massa})
for a wave propagating parallel to one interferometer arm, or its
generalization to angular dependences, the correct theory of gravity
will be massive Scalar - Tensor Gravity which is equivalent to $f(R)$
theories. In any case, such response functions will permit, in an
ultimate way, to understand if Einstein's GR is the correct and definitive
theory of gravity. This is because GR is the only modern gravity theory
which admits only the two response functions (\ref{eq: risposta totale Virgo +})
and (\ref{eq: risposta totale Virgo per}) \cite{key-3,key-4,key-5,key-8,key-17,key-18,key-30,key-31}.
Such response functions correspond to the two {}``canonical'' polarizations
$h_{+}$ and $h_{\times}.$ Thus, if a third polarization will be
present, a third response function will be detected by GWs interferometers
and this will\underbar{ }ultimately rule out GR like the correct and
definitive theory of gravity. 

Resuming, in this review we have shown that, by assuming that advanced
projects on the detection of GWs will improve their sensitivity allowing
to perform a GWs astronomy, and obtaining accurate angular and frequency
dependent response functions of interferometers for GWs arising from
various Theories of Gravity, i.e. GR and Extended Theories of Gravity,
understanding if Einstein's GR is the correct and definitive theory
of gravity will be ultimately possible.

\textbf{Acknowledgements}

The Associazione Scientifica Galileo Galilei has to be thanked for
supporting this paper.

\end{document}